\documentclass[
 twocolumn,
]{ceurart}

\usepackage{xspace}
\usepackage{listings}
\newcommand{\ie}{i.e.,\@\xspace} 
\newcommand{\eg}{e.g.,\@\xspace} 
\usepackage{amssymb}
\usepackage{pifont}
\usepackage{vcell}
\usepackage{booktabs,array,multirow,pifont}
\newcommand{\cmark}{\ding{51}}
\newcommand{\xmark}{\ding{55}}

\usepackage{fixme}
\fxsetup{status=draft, layout=inline, theme=color}
\FXRegisterAuthor{ss}{sls}{\colorbox{Purple}{\color{White}SS}}
\FXRegisterAuthor{lr}{lre}{\colorbox{Red}{\color{White}LR}}
\FXRegisterAuthor{dm}{dma}{\colorbox{Violet}{\color{White}DM}}
\FXRegisterAuthor{gg}{ggi}{\colorbox{Periwinkle}{\color{White}GG}}

\usepackage[acronym,nomain]{glossaries}
\glsdisablehyper
\newacronym{df}{DF}{Digital Forensics}\newcommand{\df}{\gls{df}\xspace}

\begin{document}

\copyrightyear{2022}
\copyrightclause{Copyright for this paper by its authors.
  Use permitted under Creative Commons License Attribution 4.0
  International (CC BY 4.0).}

\conference{APWG-EU 2025: Tech Summit and Researchers Forum,
  June 17--18, 2025, Cagliari, CA}

\title{Improving Cybercrime Detection and Digital Forensics Investigations with Artificial Intelligence}


\author[1]{Silvia Lucia Sanna}[%
orcid=0009-0002-8269-9777,
email=silvial.sanna@unica.it]
\cormark[1]
\address[1]{University of Cagliari, Cagliari, Italy}
\address[2]{Consorzio Interuniversitario Nazionale per l’Informatica, CINI, Roma, Italy}

\author[1]{Leonardo Regano}[%
orcid=0000-0002-9259-5157,
email=leonardo.regano@unica.it]

\author[1]{Davide Maiorca}[%
orcid=0000-0003-2640-4663, email=davide.maiorca@unica.it]

\author[1,2]{Giorgio Giacinto}[%
orcid=0000-0002-5759-3017,
email=giorgio.giacinto@unica.it
]

\cortext[1]{Corresponding author.}

\begin{abstract}
According to a recent EUROPOL report, cybercrime is still recurrent in Europe, and different activities and countermeasures must be taken to limit, prevent, detect, analyze, and fight it. Cybercrime must be prevented with specific measures, tools, and techniques, for example through automated network and malware analysis. Countermeasures against cybercrime can also be improved with proper \df analysis in order to extract data from digital devices trying to retrieve information on the cybercriminals. Indeed, results obtained through a proper \df analysis can be leveraged to train cybercrime detection systems to prevent the success of similar crimes. Nowadays, some systems have started to adopt Artificial Intelligence (AI) algorithms for cyberattack detection and \df analysis improvement. However, AI can be better applied as an additional instrument in these systems to improve the detection and in the \df analysis. For this reason, we highlight how cybercrime analysis and \df procedures can take advantage of AI. On the other hand, cybercriminals can use these systems to improve their skills, bypass automatic detection, and develop advanced attack techniques. The case study we presented highlights how it is possible to integrate the use of the three popular chatbots {\tt Gemini}, {\tt Copilot} and {\tt chatGPT} to develop a Python code to encode and decoded images with steganographic technique, even though their presence is not an indicator of crime, attack or maliciousness but used by a cybercriminal as anti-forensics technique.

\end{abstract}


\maketitle

\section{Introduction}
\label{sec:intro}
Traditional crime has evolved proportionally with technological advancement. The term \textit{cybercrime} has been introduced to indicate all illegal actions (\ie crime) involving digital devices, \eg phishing, online fraud, scams, ransomware, identity theft, data exfiltration, child exploitation, cyberterrorism, and cyberbullying. According to EUROPOL\footnote{\url{https://www.europol.europa.eu/crime-areas/cybercrime}}, millions of European citizens are daily victims of cybercrimes. Recently, the European Public Prosecutor's Office (EPPO)\footnote{\url{https://www.eppo.europa.eu/en/media/news/2024-annual-report-eppo-leading-charge-against-eu-fraud?utm_source=chatgpt.com}} reports a cybercrime increase of $38$\% at the end of $2024$ with estimated damage of $24.8$ billion euros. In fact, according to these statistics, most cybercrimes involve cyberattacks such as malware and network attacks. In $2013$, the European Cybercrime Centre (EC3) established support for Member States in the cybercrime fight by coordinating investigations and providing technical expertise. In a recent report, they highlighted the need for automated tools to analyze a large amount of data and the need for specific legislation. Current automated tools are more focused on malware detection and network traffic analysis. Activities regarding investigation and analysis must support cybercrime detection after cybercrime is committed and the cyberattack is successful, arresting the cyber criminals. 

For these reasons, traditional forensics science, which examines physical evidence after a crime, has been adapted to digital devices. The term \acrfull{df} indicates the analysis of digital devices after a crime. To prove the illegal activity done by a cybercriminal, their digital devices, such as mobile devices and computers, as well as enterprise servers compromised by cyberattacks, are properly analyzed. Since $1999$, NIST and subsequently ISO, developed a standard that establishes the main pillars of the \df procedure \cite{fatah_1999_nist, iso_df}. This standard defines a \df evidence as a seized digital device involved in a crime, either as the primary actor in cybercrime or as a data source for traditional crimes. To be admissible in a court case, this evidence must remain unchanged, uncorrupted, and unmodified. For this purpose, software and hardware tools (\eg write blockers) have been developed to prevent accidental evidence tampering during the \df procedure. 
The standard also defines a \df consultant as a legally mandated expert tasked by a judge in a court case to analyze digital evidence objectively, without personal interpretation or opinions on guilt, and to present their findings in a way understandable to non-technical individuals. 

To help \df practitioners in their activities, various tools, both commercial and FOSS, have been released. For example, numerous tools have been developed to automate data acquisition and analysis, including Magnet Axiom\footnote{\url{https://www.magnetforensics.com/products/magnet-axiom/}}, Inseyets\footnote{\url{https://cellebrite.com/en/cellebrite-inseyets/}}, Oxygen\footnote{\url{https://www.oxygenforensics.com/en/}}, X-Ways Forensics\footnote{\url{https://www.x-ways.net/forensics/}}. Their last versions leverage the latest advances in Artificial Intelligence (AI) to improve the speed and efficacy of the performed analysis. For example, such tools include AI-based algorithms to retrieve specific types of data (\eg pictures, audio, chats) in the acquired evidence. Such automation allows for a more privacy-oriented analysis since only data under interest for legal prosecution is analyzed. Moreover, it helps to achieve a less shocking effect and trauma in human analysts dealing with specific cases such as pedopornography \cite{Sanchez19}. These are two of the open problems of \df. In fact, it should be beneficial to preserve user privacy even if they are cybercriminals, and at the same time, avoid the analyst's shocks over time and waste time in analyzing unuseful data. However, these algorithms may not always be robust enough to be completely relied upon by \df experts. For example, a recent study demonstrated their insufficient robustness on specific presentation attacks, like deepfake images \cite{sanna_exploring_2024}. Actual systems can lack robustness because of a lack of a proper training dataset taken from real cases but anonymized accordingly. Different organizations could improve the dataset, but the regulations should allow international collaboration easily. In fact, to the best of our knowledge as \df consultants and from the experience of other international colleagues, sometimes it is hard to share knowledge. Such collaborations can also help in anti-forensics analysis with a shared ground truth on similar and specific cases and techniques. Another problem, based on our direct experience, is that sometimes data is not appropriately acquired by law enforcement or is not analyzed to preserve data integrity. Given all these problems, we claim that AI models can help in the \df analysis. It is essential to highlight that a well done \df procedure can lead to two main results: \emph{(i)} arrest the cybercriminal; \emph{(ii)} improve the cybercrime detection systems where the results of the \df analysis can be used as the input to the cybercrime detector to prevent similar crimes.

In this paper, we analyze how the latest advancements in AI may be leveraged to help detect and investigate cybercrime. 
Furthermore, we indicate how AI-based analysis may help \df practitioners in their activities. Conversely, how AI-generated content may pose challenges not adequately overcome by existing \df tools. We also elaborate on how attackers may leverage AI techniques for illegal activities. In particular, we describe a case study on the application of general-purpose LLMs for both offensive and defensive activities. In fact, as the current AI algorithms can be used for the analysis, detection and prevention of the cybercrime, also the cybercriminal can use them to improve their skills and bypass such techniques (\eg evading the \df system with anti-forensics techniques or attacks to the AI-based forensics tools, or directly to commit a cybercrime thanks to LLMs without any digital knowledge).

The remainder of this paper is structured as follows. Section~\ref{sec:literature} describes the current literature for cybercrime detection and \df analysis. Our proposals are described in Section~\ref{sec:cybercrime} and Section~\ref{sec:df}, respectively, for using AI in cybercrime and in \df. In Section~\ref{sec:casestudy} we show the use of three popular LLMs to generate steganographed images but also code for steganography image encoding and decoding. Section~\ref{sec:conclusions} closes the paper.
\section{Related Work}
\label{sec:literature}

Recent literature on cybercrime detection primarily focuses on classifying malware and detecting cyber threats across various platforms, including traditional computers, mobile devices, and IoT/OT systems. Detection systems such as Intrusion Detection Systems (IDS) and Intrusion Detection and Prevention Systems (IDPS) leverage different techniques to identify suspicious signatures, patterns, or behaviors in both network traffic and local files. Network-level tools like firewalls and honeypots help detect anomalies, while file-level analysis targets threats like phishing emails and malware binaries. Malware analysis employs both static methods (\eg hash matching, metadata inspection, code and API analysis) and dynamic techniques (e.g., execution tracing, memory monitoring, API hooking) to reveal malicious intent \cite{madhekar_literature_2024} 
Once threats are identified, IDPS systems can trigger alerts or take automatic actions, such as blocking an IP address or modifying firewall rules.

AI models are commonly integrated into these systems to enhance detection accuracy. However, due to the risk of adversarial attacks, \ie where attackers deceive the AI into misclassifying threats, these models need to be both accurate and robust \cite{tsochev_improving_2019} 
Explainable AI (xAI) techniques are increasingly applied to interpret the model's decisions, increase analyst trust, and better understand feature importance. Since some malware can detect and evade AI-based systems or behave differently in sandbox environments, hybrid detection approaches are favored to catch more sophisticated threats.

\df techniques are applied for dynamic malware detection through memory forensics, which is valuable in uncovering obfuscated behavior or retrieving ransomware keys \cite{al-sofyani_survey_2023}. Post-mortem forensics analysis is crucial for understanding attack timelines and improving future defense (DFIR) \cite{dunsin_comprehensive_2024}. Other \df efforts aim to strengthen investigation methods, assess tool reliability \cite{sanna_exploring_2024}, and develop solutions for evidence acquisition from mobile devices without requiring root \cite{Bellizzi202120}.

On the offensive side, the research explores anti-forensics strategies attackers use to hide or manipulate data to bypass human and automated detection during forensic analysis \cite{Mohammad2024}. One notable method is steganography, which conceals data within benign-looking files, making detection difficult even to human analysts \cite{dalal_steganography_2021}. This has led to the emergence of stegomalware, where malware payloads are hidden using steganographic techniques to bypass detection \cite{caviglione_never_2022}. 
\section{Cybercrime Detection with AI}
\label{sec:cybercrime}
The first step to fight cybercrime is based on its detection. Hence, AI models have the potential to significantly enhance cybercrime detection by learning common patterns found in attacks. Training these models on specific characteristics enables a better recognition of popular threat categories. However, they must be robust against adversarial attacks and capable of adapting to new variants over time through periodic retraining, addressing spatial and temporal drifts.

Generative AI (genAI) can help augment data and create new variants of specific attacks or scenarios for better accuracy in different cybercrime scenarios. In this way, the other AI algorithm used for cybercrime detection has a more extensive set of examples on which to be trained, improving the accuracy even on new attack detection. Synthetically generated data should be supervised by human expert analysts to avoid the common hallucination problem of the genAI algorithms \cite{Brewer2024525}, \ie the AI algorithms produce an output with the requested characteristics, but it has no logic or sense. This can improve the lack of a training dataset and covering a larger variety of similar cases. Large Language Models (LLMs) can be applied to chatbots, trained appropriately by expert investigators and law enforcement, to create false virtual identities and contact cyber criminals. In this way, AI can interact with cybercriminals to have data for their capture. Hence, cybercrime analysts are not exposed to dangerous scenarios. Simultaneously, data is collected for investigation or used to improve the proper AI detection algorithm by including the gathered data in the training set.

\begin{figure}
    \centering
    \includegraphics[width=0.9\linewidth]{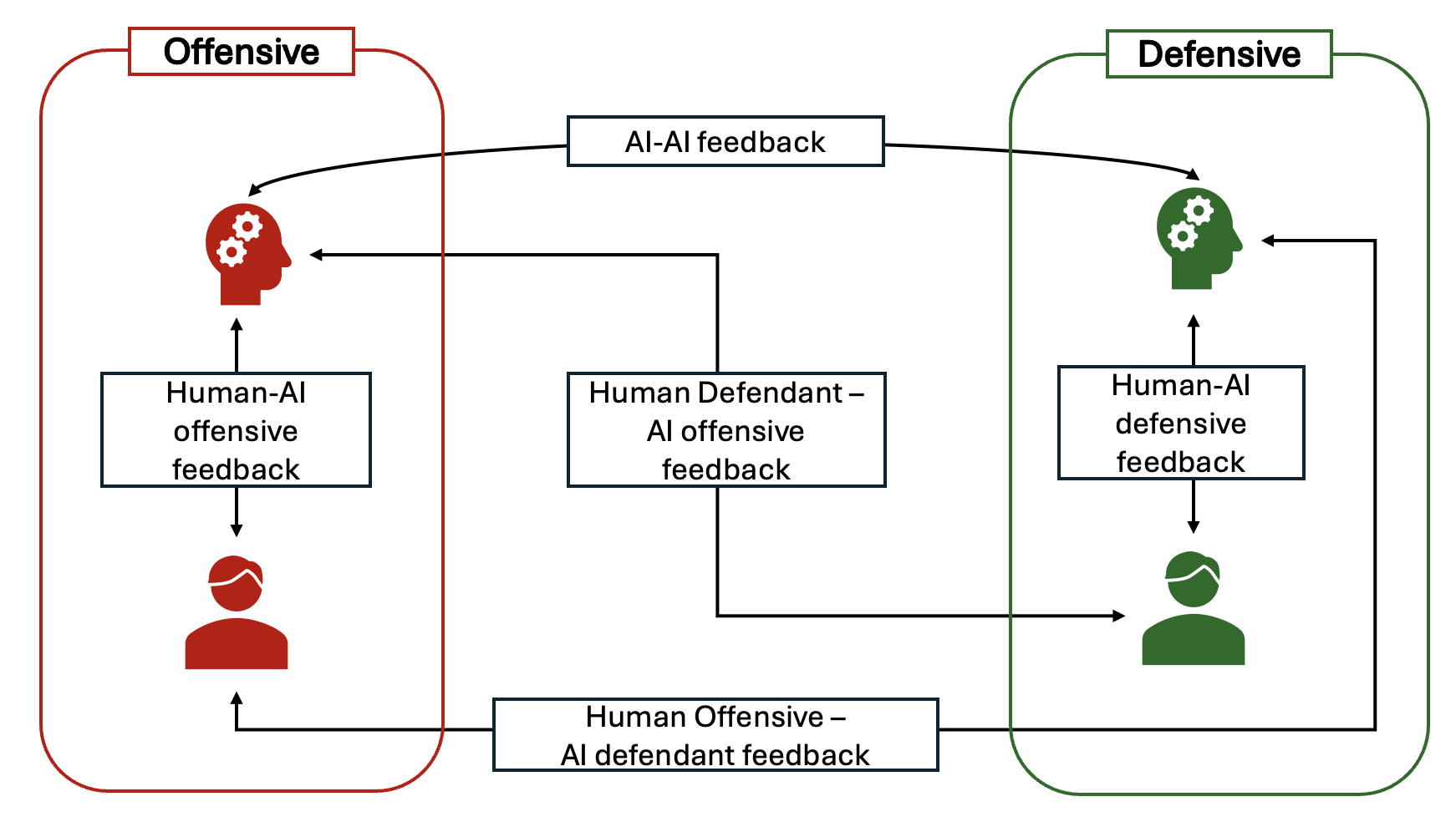}
    \caption{AI system for cybercrime detection based on two models, one for defensive (on the right, green) and one for assisting cybercriminals (on the left, red). The system is guided with human feedbacks and using AI output to retrain the counter model.}
    \label{fig:cybercrime}
\end{figure}

It is widely known that with the highest technology detection mechanisms, even with the highest accuracy and robustness, the weakest link in security is the human being. For this reason, specific training must be performed to teach the people secure behaviors to protect themselves, their data, and the infrastructure. Different studies have been published accordingly \cite{Siddiqi2022}, and psychology theories say that learning from games is the most efficient approach as concepts are assimilated while having fun\footnote{\url{https://www.brunel.ac.uk/news-and-events/news/articles/Gaming-in-the-classroom-improves-teaching-and-learning}} \cite{Shafiee20FSIDI}. A helpful exercise is the adoption of Capture-The-Flag (CTFs), \ie simulated cyber security exercises inspired by real scenarios in which people must apply real cyber security techniques to solve the problem and gain a secret word called flag to be submitted on a platform and gain points. In this case, genAI can be applied to generate new examples, exercises, and case studies for improving human learning. Additionally, an ad hoc trained AI model can be used to play against humans, as in many other popular games. An offensive AI can also be adequately trained to help specialists develop evasive methodologies and think like a cybercriminal to find a way to evade the autonomous system. The best scenario is a loop between defensive and offensive AI and with humans. In this way, the attack detection results from AI must be supervised by the human and sent to the offensive AI to develop more advanced offensive techniques, check how the human recognizes them, and this feedback is sent back to the defensive AI to improve the detection algorithm. This feedback schema is depicted in Figure~\ref{fig:cybercrime}, explaining how the system of attacking and defensive AI with human interaction helps in improving analysts' skills and AI algorithms performances. A periodic retrain must include new features, samples, and cases. 

\subsection{Ethics and Privacy}
\label{sec:cybercrime:ethics}
For more accurate crime detection, the AI algorithms can be trained on police reports and investigation information. Of course, to ensure privacy and uncorrelated data with the physical people, anonymization and appropriate security measures must be adopted (\eg protect the database with a password in a secure server with access control). Otherwise, people's data are exposed, and each person can be directly associated with crimes made or suffered. Privacy can also be preserved with AI in the analysis systems, for example, filtering specific data without examining the whole life stored in the digital device of an accused person but only the one related to the prosecution. Detection can also be improved by international collaboration between law enforcement and investigation agencies. It is essential to not have bias in the data or classify it based on racial, gender, and social origins. A more expansive global database from different parts of the world with varying background scenarios will improve the variety of training datasets, trying to avoid prejudices. To this aim, systems like shared federated learning databases, a standard AI system, and methodology must be improved. 
For this reason, a common legislative guide must be developed. Having standards for a common methodology is crucial, as cybercrime is a global issue, and law enforcement from everywhere should only fight against cybercriminals. With the incremental use of AI, specific regulations must be studied and developed accordingly to process data in a privacy-oriented manner. The algorithm must be protected by attacks not to retrieve information belonging to a specific person or easily ascribable and inferred. Moreover, if using xAI to understand the classification and improve the robustness, it must be designed not to give the person's data, \ie when explaining the classification, the xAI cannot provide as output the training sample with people's private data, which is similar to the new classified one. 

As cybercrime is based on the reasoning of the attacker, it should be interesting to study the effects of psychology. In fact, studying how the human brain reacts to attacks, \eg why people click on a phishing link, and understanding the brain reaction should help study new countermeasures. Psychology studies on the brain response could be used as feedback and additional features in developing new tools and techniques. Psychology studies can also help to understand how a cybercriminal thinks. We can train a group of researchers in attack techniques whose main objective is acting as a cybercriminal and transferring the relevant patterns to AI-based algorithms. In this way, more efficient detection techniques can be developed. 

\section{Digital Forensics with AI}
\label{sec:df}

\df is conducted after a cybercrime is perpetrated in order to arrest cybercriminals but also have a feedback on the cybercrime prevention system.
According to NIST, \df process must follow four main phases\cite{nist_guide_df}. The first one is the \textit{collection} of digital evidence, involving the identification and seizure of all devices equipped with memory, both volatile and non-volatile. Collected devices must be carefully documented to establish a \textit{chain of custody}, that is, a document recording every step of each \df procedure phase. Such devices must be preserved in their original state. If powered off, they should remain so. Otherwise, they must remain on to preserve data stored in volatile memory, isolated with appropriate measures (\eg with a Faraday bag) to prevent remote access. The second phase is the \textit{examination}, where data is acquired using appropriate techniques, depending on the type of evidence, creating a forensic copy of the original evidence. In the third phase, \textit{analysis}, investigators inspect the data on the acquired copy using specialized tools to extract relevant information guided by legal questions. 
Finally, in the \textit{reporting} phase, the expert produces a document (\ie report) summarising the case, the analysis performed and the conclusions drawn with the answers to the prosecution's questions. This document also describes the background technical concepts understandable by legal parties.

Emergent DF tools are beginning to integrate AI into their analysis systems, as described in Section~\ref{sec:intro} and ~\ref{sec:literature}, to improve the identification and search of specific media files without inspecting the whole device. This is not the only application of AI in DF. To this purpose, AI can be an additional instrument for the DF analysis and can be applied to every main phase. All the techniques presented in the following are widely used in the literature; we highlight how to apply the current methodologies to the DF pipeline. 

For the \textit{collection}, AI can help identify the digital devices to be seized at a crime scene. A camera with Computer Vision (CV) on object detection algorithms can highlight the digital devices to be seized by the first responder (\ie the police and law enforcement appointed to access the crime scene and seize them). 
This will help to not forget to seize every digital device and follow the procedure by automatically taking pictures of the found devices, identifying patterns (\eg if the device already has dents, scratches, or any damage) and reporting the data with an LLM in the chain of custody. We recommend using the CV model locally to prevent privacy without sending and collecting data on a server. Regarding \textit{examination} where the memory (\ie RAM and/or disk) of the seized digital device is acquired, AI can be applied as a digital assistant where the \df consultant writes the specifics of the case and the AI answers with a list of steps to be followed by using ad-hoc LLMs. This is very important for the reconstruction of the chain of custody. Additionally, the model is trained on different cases globally with anonymized data for better accuracy, considering only the technical procedure to guide the DF consultant. In this way, we claim that the acquisition of specific data would be facilitated by checking similarities with other cases. For example, it could help acquire corrupted memory sections or damaged disks, according to what has been done in similar cases in previous years. The \textit{analysis} phase already uses AI algorithms to classify multimedia files. Still, such algorithms must be improved in terms of accuracy and robustness to presentation attacks. AI algorithms can also be trained on specific anti-forensics and anti-analysis techniques, such as steganography, to detect hidden patterns in files or memories. Training can be improved by using a shared global database of similar anonymized cases containing different techniques to analyze and retrieve readable data. In the \textit{reporting} phase, the summary findings can be written in a draft using LLMs. For certainty, LLMs help in the background definition 
and glossary of the used terms. In the future, ad hoc trained LLMs and Natural Language Processing (NLP) algorithms can help to read the document and assign a score based on DF understandability. 

\begin{table*}[h!]
\centering
\begin{tabular}{lcccc}
    \toprule
     & \textbf{Collection} & \textbf{Examination} & \textbf{Analysis} & \textbf{Reporting} \\
    \midrule
    AI                & CV                   & LLM                    & ML \& DL             & LLM \& NLP \\
    \midrule
    Execution         & Local                & Local                  & Local                            & Local      \\
    \midrule
    Global Database          & Seized Evidence  & Data Acquisition  & Similarities & Data Correlation  \\
    \midrule
    Privacy           & Seized Scene & Personal Data & Irrelevant Data & Sensitive Information        \\
    \midrule
    Robustness        & Adversarial Attacks &    Anti-Forensics  & Presentation Attacks & Hallucination \\
    \midrule
    xAI               & Seizure & Acquisition & Classification  & Conclusions \\
    \midrule
    Human Supervision & Object Identification & Data Integrity & Misclassification & Consistency \\
    \bottomrule
\end{tabular}
\caption{Essential requirements for each DF phase. The table presents the use of specific techniques, security measures, and essential requirements for the collection, examination, analysis ,and reporting, and how to use them or for what specific action}
\label{tab:dftable}
\end{table*}

Of course, these AI models help to improve and simplify the DF workload. It is important not to rely exclusively on them, and no DF consultant could report a result just because of the detection and classification of the AI system. Supervision is needed, and AI can be seen as an additional assistant. For a better justification of results, xAI must be used to understand why the model made that specific decision, improve the robustness, and improve classification. 

\subsection{Analysis of genAI data}
\label{sec:df:genAI}
The increasing use of genAI and the easiness of generating synthetic and false data can also affect the DF world. Many tools have been developed and freely released to mimic a real person but in a fake scenario (\eg deepfakes) or ultimately generate new data from nonexistent content (\eg complete synthetic data). To the best of our knowledge, based on our personal experience in DF consultancies and feedback from other local, national, and international consultants and law enforcement, people are starting to declare in a trial that specific multimedia files (\ie audio and video) were generated with AI and they did not say or made that specific action. Hence, the main DF questions for future research on genAI data will be \textit{How to prove that the data is generated by AI mimicking the real person?} Sometimes, the human eye can recognize fake data due to recognizable artifacts in videos or pictures depicting humans. For example, such pictures may contain incoherent lights and shadows, human hands with an incorrect number of fingers, or unrealistic gestures. 
Another interesting future research question regards the attribution of the created genAI media, hence \textit{How to prove who made the synthetic audio/video/image?} This is specifically for complaint cases where fake media is generated to defame a person. In our opinion, these would be the future of DF, and many techniques must be developed accordingly. We claim that these questions could be solved with the use of steganography, for example, by adding a watermark in the file or artifacts in the metadata to trace signatures from the tool used. Some companies cannot agree on these specifics. Hence, there should be regulated standards as in the European AI Act\footnote{\url{https://digital-strategy.ec.europa.eu/en/policies/regulatory-framework-ai}}. Another strategy for detecting genAI multimedia files would be a deep file structure analysis. We know that the structure of images, audio and video follow specific patterns, \eg pixels have specific structures, audio have specific frequencies, video have specific time sequences. 
In fact, deepfakes do not always have human natural and biological signs. Recurrent unrealistic patterns, such as those highlighted previously, can be detected with a deep analysis. These patterns can also be used as features for training specific AI algorithms. Such features can also be established by a deep and detailed comparison of similar real data, by finding the differences, or by association with other similarly generated data. 

\subsection{AI helping Digital Forensics}
\label{sec:df:ai}
As described at the start of Section~\ref{sec:df}, AI can improve the DF pipeline, and current methods can be adopted during DF consultancies. In fact, the future of DF procedure can be based on AI-assisted consultancies thanks to clear patterns in similar cases worldwide. To not only rely on the AI autonomous system, the models must be used as an integration to the human analysis as a digital assistant and as a pre-analysis step where the human DF consultant can check the outcome. In particular, the pre-analysis AI-based step can help detect and extract data from challenging scenarios, such as when using anti-forensics like steganography techniques. We claim that training data can be augmented, including evidence collected from CTFs and manually validated by human experts, helping to improve detection, considering them as anti-forensics cases. In most cases, a secret message is hidden according to the author's knowledge. Sometimes, these challenges are considered guessing as people must try to understand the author's thoughts and reasoning. This procedure can be compared to analyzing the evidence belonging to a skilled person using anti-forensics techniques. In fact, in this case, the \df practitioner must try to understand what the criminal thought, for example, to hide the data. Data can be augmented in training by improving the genAI algorithm and being capable of generating synthetic data. These new synthetic data are similar to a real case but with different variants. These differences must be used as features to detect genAI data, cluster the various pieces of evidence in specific case categories, and extract artefacts more efficiently, even if anti-forensics is applied. As described in the previous paragraph, AI can recognize genAI multimedia files as currently done for deepfake detection. 
\section{Case Study}
\label{sec:casestudy}

\begin{table*}[t]
\centering
\begin{tabular}{@{}ll >{\centering\arraybackslash}p{0.8cm} >{\centering\arraybackslash}p{0.8cm} >{\centering\arraybackslash}p{0.8cm} >{\centering\arraybackslash}p{0.8cm} >{\centering\arraybackslash}p{1.5cm} >{\centering\arraybackslash}p{1.5cm}@{}}
\toprule
\multirow{2}{*}{\makebox[1.5cm][c]{\textbf{Input Image}}} &
\multirow{2}{*}{\makebox[3.5cm][c]{\textbf{Encoded String}}} &

\multicolumn{4}{c}{\textbf{Decoding Scripts}} & 
\multirow{2}{*}{\textbf{Chat Loading}} & 
\multirow{2}{*}{\textbf{Iterations}} \\
\cmidrule(lr){3-6}
& & \textbf{Gemini} & \textbf{Copilot} & \textbf{GPT} & \textbf{Zsteg} & & \\
\midrule
Gemini Generated  & $-$ & \xmark & \xmark & \xmark & \xmark & \xmark & $-$ \\
Gemini Script     & This is a secret message APWG. & \cmark & \cmark & \cmark & \cmark & \xmark & $1$ \\
Copilot Generated & This is a secret! & \cmark & \cmark & \cmark & \cmark & \xmark & $-$ \\
Copilot Script    & This is a secret message APWG. & \cmark & \cmark & \cmark & \cmark & \xmark & $1$ \\
GPT Generated     & The password is swordfish & \xmark & \xmark & \xmark & \xmark & \xmark & $-$ \\
GPT Script        & This is a secret message APWG. & \cmark & \cmark & \cmark & \cmark & \xmark & $2$ \\
GitHub Dataset    & \begin{tabular}[c]{@{}l@{}}..."rrqnDG4dja7Ga5ZdAuD77CY"\\
textView.setText(\textbackslash{}"string\_here\textbackslash{}")\end{tabular}
& \cmark & \cmark & \cmark & \cmark & \xmark & $-$ \\
\bottomrule
\end{tabular}
\caption{Decoding results for various AI-generated, scripted, and real steganographic images. The first column shows the tool to generate the input image. The second column shows the encoded string. The third column presents the decoding results for each script. The fourth column refers to the steganography extraction by loading the image in the chatbot. The last column shows the iterations needed for the correct code.}
\label{tab:decoding}
\end{table*}

Cybercriminals can exploit AI, especially general-purpose LLMs like {\tt Gemini}\footnote{\url{https://gemini.google.com/}}, {\tt Copilot}\footnote{\url{https://copilot.microsoft.com/}}, and {\tt chatGPT}\footnote{\url{https://chat.openai.com/}}, for illegal activities, including bypassing detection systems, developing anti-forensics techniques, and learning technical skills like steganography. Specifically, the use of steganography is not an indicator of crime, attack or maliciousness. Even with ethical safeguards in place, LLMs can still provide useful outputs for malicious purposes, especially when prompts are framed as part of ethical scenarios like Capture the Flag (CTF) challenges or for teaching activities. In this case study, we asked the three main LLMs to generate a steganography image with the correct decoding script. Then, we gave the image to the others to try to decode it. Notably, steganography is not an indicator of maliciousness but can be used by malware to hide a malicious payload and not be easily detected. An expert user can use steganography to hide their private content and not be easily detected by automatic tools or by the human eye. We used {\tt Gemini 2.5 Pro}, the default version of {\tt Copilot} and {\tt chatGPT-4o}. We asked the three LLMs to generate an image with steganographed content and the corresponding Python code to extract the content. This is the prompt used for the three chatbots: \textit{Can you generate an image with a hidden secret with steganography techniques? I also want the script python to decode it}. Then, we saved the images and the scripts and asked for more details from the chatbots about the encoded message and the techniques used. They all used LSB, maybe because it is the most popular and straightforward methodology. Unfortunately, none of the three LLMs could generate a PNG working image with steganographed content. We used {\tt zsteg}\footnote{\url{https://github.com/zed-0xff/zsteg}} external Linux tool to extract the steganographed content by checking the images; neither this tool could extract the secret from the generated image. Other tools like {\tt steghide} and {\tt stegcracker} are developed for other image formats not PNG \eg jpg, bitmap. Neither of their decoding scripts could extract meaningful content. For this reason, we asked to generate a script on an input PNG image and embed a given secret, saving the output image. Even in this part, all of them chose the LSB technique. We gave as input a white png image taken from Wikimedia\footnote{\url{https://upload.wikimedia.org/wikipedia/commons/d/d2/Solid_white.png?20060513000852}}, embed the secret \textit{This is a secret message APWG.}, and saved the steganographed content. Then, we used their generated decoding script, and in all the cases, they could extract the steganographed content. We loaded these images in the chatbots, but none of the images could be decoded by the LLMs, not the one generated by the same script nor the one generated by the other LLMs. The generated scripts are not similar. Only with {\tt Copilot} we had to clarify it was for teaching activities, otherwise it has been detected as a malicious prompt. In the GitHub page\footnote{\url{https://github.com/slsanna/LLMs-Chatbots-for-Steganography}} with the chats, the generated scripts, and images, but only with the chatGPT decoding script we had to re-iterate it because of a bug in the computation of the pixels to encode the string. To prove the generated scripts' efficiency, we tested the decoding on a dataset\footnote{\url{https://github.com/Ocram95/AndroidStego/tree/main/resources/stego_resources/LSB/Sequential}}\footnote{\url{https://github.com/Diesoi/AndoidStegoLoader/tree/main}} and all the three generated decoding scripts could extract the steganographed string. The summary results reported in~\ref{tab:decoding} show that none of the tools can generate a correct steganographed png image, but the script generated for encoding/decoding generally works from the first output (\ie the first given answer by the chatbot), they are correct also in a real dataset, but the chatbot itself cannot extract the content from the image.

\section{Conclusions}
\label{sec:conclusions}
This paper analyzes the current literature on using AI in cybercrime detection and \df. Because the literature is not exhaustive, we present some future scenarios where we claim AI can be adopted in cybercrime detection and DF analysis. We stress the use of AI detection systems on popular and common cybercrimes, using LLMs for chatbots, and the importance of ethics, privacy, and psychology. About \df, we designed how to apply AI in each \df stage and how to determine the attribution of media to AI-generated data forensically. In particular, we presented a little case study where the popular LLMs chatbots {\tt Gemini}, {\tt Copilot} and {\tt chatGPT} can assist both case analysis and in an offensive way. In fact, we tested their performance in generating images with steganographic content using LSB and their capability to generate code for encoding and decoding. The most performant LLM for this purpose is {\tt Copilot}. We also tested their generated data in an external dataset available on GitHub and the Linux tool {\tt zsteg}. Future studies must be conducted to test their performance better in other analysis and anti-forensics scenarios and also consider the development of ad-hoc tools. We claim that in the future, AI will be used more exhaustively in cybercrime and digital forensics, up to a point where there will be a system of entirely autonomous AI-vs-AI for defense and attack. 
\section*{Acknowledgments}

This work was partially supported by Project SERICS (PE00000014) under the NRRP MUR program funded by the EU - NGEU. This work was carried out while Silvia Lucia Sanna was enrolled in the Italian National Doctorate on Artificial Intelligence run by Sapienza University of Rome in collaboration with the University of Cagliari.
\bibliography{sample-ceur}
\end{document}